\documentclass[12pt]{iopart}
\usepackage{graphicx}
\usepackage{iopams}
\begin{document}
\title[A new method ...]{A method to quantitatively evaluate Hamaker constant using the 
jump-into-contact effect in Atomic Force microscopy}
\author{Soma Das, P. A. Sreeram and A. K. Raychaudhuri}
\address{DST Unit for Nanoscience.\\
S. N. Bose National Centre for Basic Sciences,\\
Block JD, Sector III, Kolkata 700098, India.}
\eads{\mailto{soma@bose.res.in}, \mailto{sreeram@bose.res.in}, \mailto{arup@bose.res.in}}
\begin{abstract}
We find that the ``jump-into-contact'' of the cantilever in the Atomic Force Microscope (AFM) is caused by an inherent instability in the motion of the AFM cantilever. The analysis is based on a simple model of the cantilever moving in a non-linear force field. We show that  the ``jump-into-contact'' distance can be used to find the interaction of the cantilever tip with the surface. In the specific context of the attractive van-der-Waals interaction, this method can be realized as a new method of measuring the Hamaker constant for materials. The Hamaker constant is determined from the deflection of the cantilever at the ``jump-into-contact'' using the force constant of the cantilever and the tip radius of curvature, all of which can be obtained by measurements. The results have been verified experimentally on a sample of cleaved mica, a sample of Si wafer with natural oxide and a silver film, using a number of cantilevers of different spring constants. We emphasize that the method described here is applicable only to surfaces that have van-der-Waals interaction as the tip-sample interaction. We also find that the tip to sample separation at the ``jump-into-contact'' is simply related to the cantilever deflection at this point and this provides a method to exactly locate the surface.
\end{abstract}
\maketitle

\section{Introduction}

\begin{figure}[h]
\vspace{0.5cm}
\centerline{\includegraphics[height=5cm]{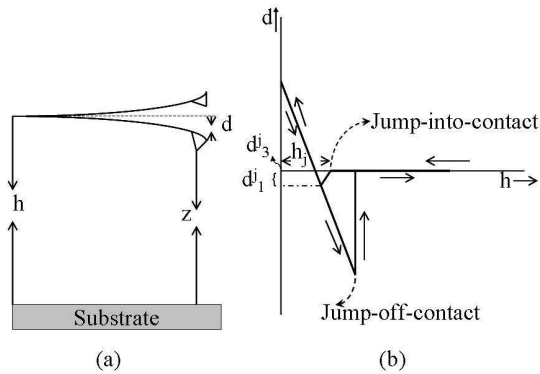}}
\caption{Schematic diagram of AFM tip and sample assembly (a) and force-distance curves (b). The dotted line in (a) marks the equilibrium position of the cantilever in absence of an external force. ``d'' is positive when measured upward. The arrows in (b) show the direction of motion of the cantilever.}
\label{afm}
\end{figure}

Atomic Force Microscope (AFM) is one of the most widely used tools in nanoscience and nanotechnology.
Since its discovery, AFM \cite{binning86} has emerged as a very powerful tool in the characterization of various properties of materials at the nanometer scale. This is primarily because AFM can not only image with atomic resolution but it can also measure interatomic forces which are of the order of pico Newtons or even much less. These capabilities made AFM a versatile enabling tool in nanotechnology. One of the standard experiments performed by AFM  is the measurement of the force-distance curves \cite{cappella99,butt05} which measures the force of interaction between the tip and the substrate. In this measurement, the cantilever deflection ($d$) is measured as a function of the separation of the tip and the sample ($z$) and the force of interaction is the product of the deflection `$d$' of the
cantilever and the spring constant `$k_c$' of the cantilever. (Note: The force obtained in this manner is not exactly the force between the tip and the sample, since the 
effective spring constant of the cantilever can be modified by the elastic deformation of the surface of the sample and the tip when they are in contact with each other. Hence, in our study, we will consistently use the concept of deflection of the cantilever instead of the force.).
In the  measurement of the force-distance curve, $d$ is measured from its equilibrium position (in the absence of any external force), when it is at a distance `$h$' from the sample (the substrate) as shown in \fref{afm}a. 
The measured force-distance curve, shown schematically in \fref{afm}b generally shows hysteresis. The approach curve shows a ``jump-into-contact'' (JC) and the retraction part of the curve shows the ``jump-off-contact''. The concept of ``jump-off-contact'' has been used extensively in the past as a quantitative measure of the adhesion force \cite{capella97,hao91}. One of the most ubiquitous explanations for the ``hysteresis-like'' behaviour observed in the force-distance curves in AFM is based on the presence of adhesion forces, due to a layer of water on the surface of the sample and the tip \cite{sarid98}.  In contrast, however not much attention has been paid to the phenomena of ``jump-into-contact'' (JC), except early papers that pointed out the basic causes for existence of such a phenomena \cite{cappella99,colton89}. In this paper, we revisit the issue of ``jump-into-contact'' again and present a new approach to ``jump-into-contact'' phenomenon in the force-distance curves of AFM. This is done in order to investigate whether it can be used to obtain quantitatively some of the microscopic parameters of tip-sample interaction and thus can be made an useful tool. We find that JC is a generic manifestation of the  fundamental instability in the motion of the cantilever in a non-linear force field. A simple model is used to understand this instability and obtain a quantitative measure of not only the distance `$h_j$' at which the JC should occur but also how much should be the magnitude of the deflection of the cantilever at the JC.  These measures are directly related to parameters of the force field. We have performed experiments to verify some of the predictions of our theory. In this paper we investigate this phenomena in the specific context of the van-der-Waals interaction and from the measured deflection of the cantilever at the JC we determined the Hamaker constant using the known parameters like the radius of curvature of the tip ($R_t$) and the cantilever spring constant ($k_c$). In addition, our investigation also gives us a practical way to locate the actual distance of the cantilever from the surface. The phenomena is completely governed by the attractive part of the tip-sample interaction. An important outcome of the investigation is the observation of the independence of the JC on the actual elastic forces that make the tip-sample contact interaction. Since the JC can be measured at different spots on a given surface, this method gives us a tool to obtain a map of the tip-surface interaction as measured by parameter like Hamaker constant (in the context of van-der-Waals interaction) with a spatial resolution offered by a typical AFM. The spatial resolution is an added advantage over other methods of determining the Hamaker constant like the surface force apparatus.

The paper is organized as follows. In Section 2, we introduce our simple spring-ball model for the motion of the  cantilever. We solve the static force equation analytically to locate the instability that makes the JC to happen. In Section 3, we show experimental data of force-distance curves to measure the JC distance, experimentally verify some of the theoretical predictions and obtain the Hamaker constant. We also discuss the extent of uncertainty in the data and compare the relative merits of this method vis-a-vis other methods of determining Hamaker constant. In section 4, we present our conclusions.

\section{Theoretical Modelling}
The AFM is a  nonlinear system. Our aim here is to use a simple model which could explain the feature seen in experiments. We model the motion of a cantilever by a spring-ball system. The inherent nonlinearity of the cantilever due to its finite dimensions have not been introduced into our calculation, in order to keep things simple. Thus, we write
the equation of motion of the cantilever as,
\begin{equation}
m \ddot d(t) + \eta \dot d(t) + k d(t) = f_{ts}(h+d(t)).
\label{dynamic_eqn}
\end{equation}
Here, $m$ is the mass of the cantilever, $\eta$ is the friction constant, $k_c$ is
the spring constant, $d(t)$ is the deflection of the cantilever measured from its
equilibrium position in the absence of any external force, $h$ is the distance
between the sample and the tip when the tip is in the equilibrium position (in
the absence of any external force) and $f_{ts}(h+d)$ is the atomic force between
the tip and the sample at the instantaneous position of the tip and $t$ represents time. In case of the static (or quasi-equilibrium) experiment $d(t)=d$, where $d$ is the deflection of the cantilever at which it comes to rest. The dynamic equation  will reduce to a simple static equation of the form,
\begin{equation}
k_c d = f_{ts}(h+d),
\label{static_eqn}
\end{equation}

One can take a generalized force field for $f_{ts}(h+d)$ and obtain a solution to the  \Eref{static_eqn} that will give the parameters of the interaction potential. In order to have a definite result that can be verified by experiment, we investigated the specific case of van-der-Waals interaction between the tip and the surface and an elastically deformable surface for the contact force. The subsequent results obtained are thus specific to the van-der-Waals interactions. The tip-sample force is modelled by a combination of van-der Waals force at large tip-sample distances ($h$) which is essentially attractive and by the Dejarguin-Muller-Toporov(DMT) \cite{israelachvili91,derjaguin75} force which is a combination of the attractive van-der-Waals-like force (except that it is $h$-independent) and the repulsive forces arising due to elastic interaction between the tip and the sample. Thus, formally, the force is given by,
\begin{equation}
\label{force_eqn}
f_{ts}(z) = 
\cases{ - \frac{H R_t}{6 z^2} & for $z>a_0$,
\\
 - \frac{H R_t}{6 a_0^2} + \frac{4}{3}E^\star \sqrt{R_t} (a_0-z)^{3/2} & for
$z\leq
 a_0$.\\
}
\end{equation}
Here, $z=h+d$, $a_0$ is an intermolecular distance, $H$ is the Hamaker constant, which depends on the material of the tip and the sample and also on the intervening medium. $E^\star$ is the effective elastic modulus between the tip and the sample. 
%\begin{equation}
%\frac{1}{E^\star}=\frac{(1-{\nu_t}^2)}{E_t}+\frac{(1-{\nu_s}^2)}{E_s}
%\end{equation}
Note that the form of the van-der-Waals force is chosen for a sphere-plate geometry, which is close to the real situation in an AFM experiment. In this paper, we will only concentrate on the regime where $z > a_0$, where 
the force is purely a van-der-Waals force. We will see below that the JC is mainly determined by the attractive part of the interaction.

For the observation of JC, we work in the region of attractive interaction and 
 take the force on the right hand side of \Eref{static_eqn}
to be the van-der-Waals force. This is justified because, we will see below that the ``jump-into-contact'' distance is usually much larger than $a_0$. From Equations \eref{static_eqn} and \eref{force_eqn}
 after some simple manipulations, we obtain the
 equation for the deflection (d) as,
\begin{equation}
k_c d (d+h)^2+\frac{HR_t}{6}=0.
\label{static1_eqn}
\end{equation}
Rewriting $\tilde d$ = $d/h$ and $\tilde a$=$HR_t/6k_{c}h^3$, we get,
\begin{equation}
\tilde d(1+\tilde d)^2+\tilde a = 0.
\label{mod_static}
\end{equation}
The three solutions of this equation are given by,
\begin{eqnarray}
\tilde d_1 &=& -\frac{b_2}{3} + (S+T) \nonumber \\
\tilde d_2 &=& -\frac{b_2}{3} -\frac{1}{2}(S+T)+\frac{\sqrt{3}}{2}\imath(S-T)
\nonumber \\
\tilde d_3 &=& -\frac{b_2}{3} -\frac{1}{2}(S+T)-\frac{\sqrt{3}}{2}\imath(S-T)
\label{solutions}
\end{eqnarray}
where,
\begin{eqnarray}
S = \left(R + \sqrt{D}\right)^{1/3} \nonumber \\
T = \left(R - \sqrt{D}\right)^{1/3} \nonumber \\
R = \frac{9b_1b_2-27b_0-2b_2^3}{54} \nonumber \\
D = Q^3 + R^2 \nonumber \\
Q = \frac{3b_1-b_2^2}{9}
\label{components}
\end{eqnarray}

\begin{figure}[h]
\vspace{0.5cm}
\centerline{\includegraphics[height=5cm]{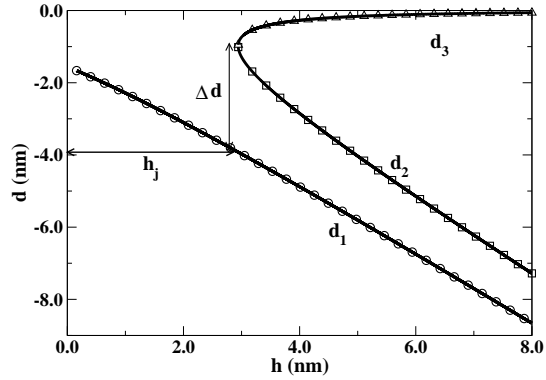}}
\caption{Plot of solutions given by \Eref{solutions} as a function of
tip-sample distance (h) for the parameters mentioned in the text. The open circles ($d_1$=$\tilde d_1.h$) and open triangles ($d_3$=$\tilde d_3.h)$ represent stable solutions. The open squares ($d_2$=$\tilde d_2.h$) represent the unstable solution. Here only the real part of the solutions in shown.}
\label{sol_fig}
\end{figure}
and $b_0$, $b_1$ and $b_2$ are the coefficients of $(\tilde d)^0$, $(\tilde d)^1$
and $(\tilde d)^2$ in \Eref{mod_static}. Here we want to mention that \Eref{solutions} is valid only for the real values of S and T defined in \Eref{solutions}. For complex values, the expressions for \Eref{solutions} will change. It can be easily seen that, 
$b_0$ = $\tilde a$, $b_1$ = 1 and $b_2$ = 2. The distance $\tilde d_1$ has only a real
part, while the solutions $\tilde d_2$ and $\tilde d_3$ are either both real
or complex conjugate of each other, depending on the parameters of the equation.
The actual deflection ($d$) is obtained by multiplying the solution by the
corresponding tip-sample distance ($h$). \Fref{sol_fig} shows the
solutions of the actual deflection ($d$) as a function of the tip-sample distance ($h$)
for $HR_t$=2.26 x 10$^{-27}$ N.m$^2$ and $a_0$=0.15 nm. Of the three solutions, the solution given by the open circles corresponds to $d_1$ = $\tilde d_1.h$ while the open square and triangle correspond to $d_2$ = $\tilde d_2.h$ and $d_3$ = $\tilde d_3.h$ respectively. Note that as the tip-sample distance is reduced, the solutions corresponding to $d_2$ and $d_3$ approach each other and they meet at one point (for example, at $h$ $\approx$ 2.9 nm in \fref{sol_fig}). For tip sample distances below this both these solutions become complex (in \fref{sol_fig} the real part only is shown). It is necessary to note that the solutions
$d_1$ and $d_3$ are stable solutions, while $d_2$ is unstable. This has been checked by finding the sign of the derivative of  \Eref{static_eqn} with respect to $d$ at each value of $h$. We denote the point where the solutions $d_2$ and $d_3$ meet as the ``jump-into-contact'' point. This is the limit of stability for the solution  $d_3$ which defines the motion of the cantilever for the approach curve till this point.  If the tip-sample distance ($h$) is reduced beyond this point of stability,  there is only one real solution available ($d_1$) and the system will jump into the stable solution given by $d_1$. This defines the ``jump-into-contact''. It must be noted here that this jump has occurred in the attractive regime and we do not take recourse to any adhesion forces for explaining the phenomena. We also emphasize here that on the retract path the cantilever dynamics follow the solution given by $d_1$ until it jumps back to the solution given by $d_3$ at the ``jump-off-contact'' point. 
The solutions of the cubic equation given by \Eref{solutions} have a number
of interesting features. For example, let us consider the point where the
``jump-into-contact'' occurs in our model. At this point $d_2$ = $d_3$ and the discriminant $D$ is exactly equal to zero. If we denote the tip-sample distance at this point by `$h_j$' and corresponding values of $R$, $Q$ as $R_j$, $Q_j$, we get the equation,
\begin{equation}
 R_j+\sqrt{Q_j^3+R_j^2}=R_j-\sqrt{Q_j^3+R_j^2}
\end{equation}
 which leads to the equation,
\begin{equation}
 Q_j^3=-R_j^2
\end{equation}
Replacing the expressions for $Q_j$ and $R_j$ from \Eref{components}, and
putting in the values of $b_0$, $b_1$ and $b_2$, we get,
\begin{equation}
 \frac{HR_t}{6 k_c h_j^3}=\frac{4}{27}
\label{ham-1}
\end{equation}

\Eref{ham-1} can be used to find the Hamaker constant ($H$) because the tip radius $R_t$ and the cantilever spring constant $k_c$ are known and $h_j$ is experimentally measureable. However, the problem arises because the postion of the surface being not known exactly, the absolute value of $h_j$ has a large uncertainty. Below we show that the magnitude of the jump of the cantilever at JC is simply related to $h_j$ and this fact can be used to calculate the Hamaker constant ($H$) with high degree of confidence which is limited by the magnitude of the uncertainty in determination of $k_c$ and $R_t$ both of which, however, are experimentally measurable \cite{cook06,charles05}. 

At the JC, there are only two distinct real solutions to the cubic equation since the solutions corresponding to $d_2$ and $d_3$ are degenerate. Subtracting $d_3$ from $d_1$, and again putting in the values of $b_0$, $b_1$ and $b_2$, we get the jump of the cantilever ($\Delta d$) at JC as,
\begin{equation}
 \Delta d=d^j_3-d^j_1=-h_j
\label{ham-2}
\end{equation}
where $d^j_1$ and $d^j_3$ are the deflections at the ``jump-into-contact'' point corresponding to the two solutions. 
Equations \eref{ham-1} and \eref{ham-2} lead to a practical  way of calculating the Hamaker constant from the deflection-displacement curves. Determination of Hamaker constant from equations \eref{ham-1} and \eref{ham-2} will need knowledge of $k_c$ and $R_t$ of a given cantilever which can be obtained from experiment. Alternatively, we note that if $h_j$ is measured for a material with known Hamaker constant, this can be used to calibrate ($k_c/R_t$) ratio of a given cantilever, which in turn can be used to find an unknown Hamaker constant. Given the practical difficulties in knowing $k_c$ and $R_t$ exactly this may be a more practical method. Note also that \Eref{ham-2} is
itself independent of the material of the tip, sample and the intervening medium. The above mentioned process also indicates that one can obtain a precise method of shifting the raw data obtained from the AFM measurements to properly locate the surface.

\section{Experimental verification} 
The data has been taken using an Atomic force microscope (Model CP II) from Veeco \cite{Veeco} on freshly cleaved mica, on Si wafer with natural oxide and on a silver metal film. We have used three different cantilevers for taking data for a given surface in order to vary ($k_c/R_t$) ratio. The cantilevers used had  Si$_3$N$_4$ tips and spring constants (radius of curvature) of 0.03 N/m (R$_t$ = 30 nm), 0.1 N/m (R$_t$ = 35 nm) and 0.9 N/m (R$_t$=50 nm). We have found out the radius of curvature of the tip from the  images  taken by a Field Emission Gun Scanning Electron Microscope (FEG-SEM). The medium between the tip and sample for all the experiments was air at room temperature and the rates of data collection were 0.5 Hz and 0.1 Hz. We have repeatedly taken the force-distance curves using the same three cantilevers mentioned above. The reproducibility of the data confirms that there was no damage of the tip of the cantilever during the experiment and this is also corroborated by FEG-SEM image.

\begin{figure}[h]
\vspace{0.5cm}
\centerline{\includegraphics[height=5cm]{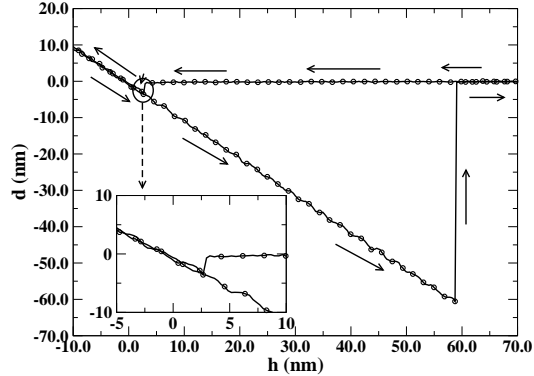}}
\caption{Approach and retract curves of the deflection ($d$) versus the
displacement ($h$) of the tip of the microcantilever for Mica using a
Si$_3$N$_4$ tip (free motion spring constant k=0.1 N/m) in air. The arrows indicate the direction of motion of the cantilever (approach and retract). The JC region is highlighted in the inset.}
\label{air_full}
\end{figure}

\Fref{air_full} shows a typical AFM deflection ($d$) versus displacement ($h$)
curve for a freshly cleaved Mica sample and a Si$_3$N$_4$ tip of spring constant 0.1 N/m in air. The data has been plotted as deflection versus distance. The arrows in \fref{air_full} indicate the direction of motion of the cantilever (approach and retract). The JC region is highlighted in the inset of \fref{air_full}. From \fref{air_full} we can see that the ``jump-into-contact'' occurs at tip-sample separation ($h$) of approximately 2.9 nm which is the attractive regime of the force distance curves, since $a_0\approx 0.15$ nm \cite{israelachvili91,derjaguin75}. The ``jump-into-contact'' occurs at larger values of $h$ for smaller $k_c$. For instance, for $k_c\approx 0.03$ N/m, the jump occurs at $h_j\approx 3.5$ nm. Thus the parameters of the attractive potential are good enough to determine the JC. 

\begin{figure}[h]
\vspace{0.5cm}
\centerline{\includegraphics[height=5cm]{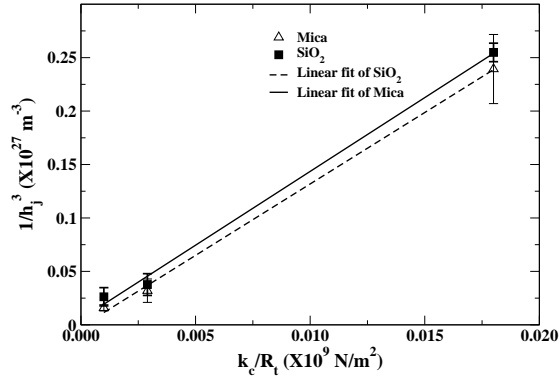}}
\caption{Plot of the observed jump into contact distance $h_j^{-3}$ as a function of the quantity $k_{c}/R_{t}$ . The solid line and the dashed line are the best fits of the experimental data on mica and Si wafer with natural oxide respectively.}
\label{jcd}
\end{figure}

In \fref{jcd} we plot the observed $h_j^{-3}$ ($h_j$ is jump into contact distance) as a function of the quantity ($k_{c}/R_{t}$) ratio. The ratio ($k_c/R_t$) are the physical parameters of the cantilevers used. We have taken three cantilevers of same composition but with different $k_c$ and $R_t$ to achieve three different ($k_c/R_t$) ratio for a given surface material. The main uncertainty in determination of $H$ from the experiment arises from these two parameters. One can use the parameters given by the manufacturer's data but a better alternative is to experimentally determine them. From \Eref{ham-1}, it can be seen that since the $\Delta d$ or $h_j$ $\propto$ $k_c^{-1/3}$, thus it is advisable to use a softer cantilever (low $k_c$) so that $\Delta d$ or $h_j$ are larger leading to less uncertainty in determination of these quantities. The graph according to Equations \eref{ham-1} and \eref{ham-2} give a straight line and the inverse of the slope gives the Hamaker constant $H$. The data have been taken in ambient within a  glove chamber. The error bar in the data have been obtained by repeated data taking on the same surface and with the same tip and it shows the extent of variance one would expect in such experiments. The reproducibility of the data also indicates that the tip, used in the experiment, did not get damaged during the collection of the data. From our experiment we obtain $H$ $\approx$ (0.64  $\pm$ 0.07) $\times$ 10$^{-19}$ J for mica, $H$ $\approx$ (0.66 $\pm$ 0.27) $\times$ 10$^{-19}$ J for SiO$_2$ and $H$ $\approx$ (3.73 $\pm$ 0.89) $\times$ 10$^{-19}$ J for silver. A summary of results obtained is shown in Table 1. This can be compared with calculated values of 1.28  $\times$  10$^{-19}$ J for mica \cite{lennart97},  1.21 $\times$  10$^{-19}$ J for SiO$_2$ \cite{lennart97} and 2.9 $\times$  10$^{-19}$ J for silver \cite{parker92}, using Si$_3$N$_4$ as the tip material with vacuum as the intervening medium . A similar calculation with water as the intervening medium gives 0.245 $\times$  10$^{-19}$ J for Mica \cite{lennart97}, 0.207 $\times$  10$^{-19}$ J for SiO$_2$ \cite{lennart97} and 1.39 $\times$  10$^{-19}$ J for silver \cite{visser76} with Si$_3$N$_4$ as the tip material . These values have been obtained from calculations using full Lifshitz theory \cite{lifshitz56,dzy61,ninham70} for the individual materials. The values we have obtained experimentally lie between the Hamaker constant values for vacuum as the intervening medium and water as the intervening medium, suggesting the influence of relative humidity of air in calculating the Hamaker constant. The relative humidity in the glove chamber during measurement was typically $\sim 55\%$ (for mica and SiO$_2$) and $\sim 33\%$ for silver. Since the JC data are routinely obtained while one takes the force-distance curve it is easy to obtain a very important physical parameter from the same experiment. We point out that the utility of JC data to obtain a quantitative measure of the physical parameter like the Hamaker constant is novel and the uncertainty in determination of $h_j$ can be eliminated by measurement of $\Delta d$ and the ($k_c/R_t$) ratio. As mentioned before, a known tip-surface system (known Hamaker constant) can be used to calibrate a given cantilever ($k_c/R_t$) ratio using Equations \eref{ham-1} and \eref{ham-2} if no direct measurement of $k_c$ and $R_t$ are available. This calibration can also be used to find Hamaker constant for unknown surface. We also emphasize that since for a given cantilever, ($k_c/R_t$) ratio is fixed a map of JC on an inhomogeneous surface can generate a map of Hamaker constant. 

There are quite a few methods of measuring Hamaker constant \cite{harold96}. These methods include direct force measurements using surface force apparatus \cite{horn88} and atomic force microscope \cite{butt05,nancy90,ederth01,seog02} where the full force-distance curve is fitted to a model of the van-der-Waals equation. The other methods are based on measuring physical properties of materials like dielectric constant \cite{tabor69,hough80}. In general, Lifshitz theory \cite{lifshitz56,dzy61,ninham70} is widely used for calculating Hamaker constants from the dielectric constant of materials. We note that, the reported values of Hamaker constants for the same material obtained from different methods showed considerable variations \cite{harold96}. The earlier works in obtaining Hamaker constant from Atomic force microscope were mainly based on fitting the attractive part of the approach curve with the expression for the van der-Waals force. In that method the main problem was the presence of the ``jump-into-contact''. In our method we have actually used the ``jump-into-contact'' to find the interaction constant. 
%\begin{widetext}
\begin{table}
\begin{center}
\begin{tabular}{|c|c|c|c|}
\hline
Materials & Atomic Force Microscope & Surface Force Apparatus & Full Lifshitz Theory \\
  ~ & (using ``jump-into-contact'') (10$^{-19}$ J) & (10$^{-19}$ J) & (10$^{-19}$ J) \\
\hline 
Mica & 0.65 $\pm$ 0.07 & 1.35, 0.22 \cite{harold96} & 1.28, 0.245 \cite{lennart97} \\
      & (air, $R_H$ $\sim 55\%$) & (vacuum), (water) & (vacuum), (water) \\
\hline
Si$0_2$ & 0.62 $\pm$ 0.27 & 50-60 \cite{harold96} & 1.21, 0.207 \cite{lennart97}\\
  ~ & (air, $R_H$ $\sim 55\%$) & (vacuum) & (vacuum), (water) \\
\hline
%Si$0_2$ & --- & --- & 0.207 \cite{lennart97} \\
 % ~ & (water) & (water) & (water) \\
%\hline
Silver & 3.73 $\pm$ 0.89 & 2.9 \cite{parker92} & 2.4 \cite{coakley}, 1.39 \cite{visser76}\\
  ~ & (air, $R_H$ $\sim 33\%$) & (air) & (vacuum), (water) \\
\hline
\end{tabular}
\caption{Values of Hamaker Constant obtained by our method and its comparison with Lifshitz theory. We also show experimental values obtained by surface force apparatus.} 
\end{center}
\end{table}
%\end{widetext}

The advantage of measuring Hamaker constant using the method described in this paper is that we can find the ``jump-into-contact'' distance from experimental force-distance curves using \Eref{ham-2} easily. No numerical fit to the complete force-distance curve is necessary. Here we also want to emphasize that this method has the advantage of mapping the Hamaker constant in an inhomogeneous system whereas it is not possible using surface force apparatus because it does not have the spatial resolution. Two important points have to be noted in this context - first, the experimental force-distance curves should be taken properly for approach  of the cantilever motion with close measurements near the JC and second, the radius of the tip ($R_t$) and the spring constant of the cantilever ($k_c$) have to be found out with least uncertainty if absolute data have to be obtained. 

We point out that \Eref{ham-2} is also a  very important outcome of this work. This gives us a way to determine the actual position of the surface. In AFM measurements there is indeed a problem in evaluating the absolute value of the tip-sample separation ($h$). The JC is a special point at which the distance ($h$) is equal to $\vert{\Delta d}\vert$ , which thus can be appropriately fixed. Once this is fixed the position of origin of $h$ (the sample) can be located. We emphasize that the analysis above is applicable only to surfaces that have van-der-Waals interaction as the tip-sample interaction because of the specific type of tip-sample interaction used. However, the method is general enough and can be used with other tip-sample interaction as well. The fact that nonlinear force field introduces an instability that leads to ``jump-into-contact" is a conclusion of general validity.

\section{Conclusions}
 We have studied the static deflection-distance curves for an atomic force
microscope using a simple model that gives ``jump-into-contact''often observed in force-distance curves of AFM as an instability of the cantilever moving in a non-linear force field. The model provides a unique method of determining the tip-sample interaction parameters. We have developed the concept specially for van-der-Waals interaction for definiteness. In this case the method gives the Hamaker constant. We find values that are comparable to the Hamaker constant measured by other methods. The model also provides a reliable criterion for locating the sample and thus  shifting the raw deflection-distance data obtained from AFM by locating the distance at which the``jump-into-contact'' occurs. This process removes the arbitrariness of locating the sample in AFM. The method also gives us a way to map the Hamaker constant over a surface, that may be inhomogeneous, by mapping the JC with AFM. 

\ack{}
The authors want to thank the Department of Science and Technology, Government of India for financial support as a Unit for Nanoscience. One of the authors, (PAS) would like to thank  Prof. B. Dutta Roy for very useful suggestions.

\end{document}